\documentclass[runningheads,a4paper]{llncs}

\usepackage{microtype} 
\usepackage{amssymb,amsmath}
\usepackage{xspace}
\usepackage{color}
\usepackage{graphicx}
\usepackage{hyperref}
\usepackage{subcaption}
\usepackage{wrapfig}

\newcommand{\BeginMyItemize}{\begin{itemize}\setlength{\itemsep}{-\parskip}}
\newcommand{\EndMyItemize}{\end{itemize}}
\newcommand{\myitemize}[1]{\BeginMyItemize #1 \EndMyItemize}

\newcommand{\BeginMyEnumerate}{\begin{enumerate}\setlength{\itemsep}{-\parskip}}
\newcommand{\EndMyEnumerate}{\end{enumerate}}
\newcommand{\myenumerate}[1]{\BeginMyEnumerate #1 \EndMyEnumerate}

\newtheorem{defin}{Definition}
  
\newtheorem{theo}[defin]{Theorem}
  
\newtheorem{lem}[defin]{Lemma}

\newtheorem{propo}[defin]{Proposition}
  
\newtheorem{cla}[defin]{Claim}
  
\newtheorem{coro}[defin]{Corollary}
  
\newtheorem{obse}[defin]{Observation}

\newenvironment{myproof}{\emph{Proof. }}{\hfill $\Box$ \medskip\\}

\renewcommand{\paragraph}[1]{\vspace*{2mm}\noindent{\bf #1}}

\newif\ifComments
\Commentstrue
\ifComments
  \newcommand{\mdb}[1]{\textcolor{blue}{MdB: #1}}
  \newcommand{\am}[1]{\textcolor{red}{AM: #1}}
\else
  \newcommand{\mdb}[1]{}
  \newcommand{\am}[1]{}
\fi

\newif\ifOmit
\Omittrue

\DeclareMathOperator{\polylog}{polylog}

\renewcommand{\leq}{\leqslant}
\renewcommand{\geq}{\geqslant}

\newcommand{\Reals}{{\Bbb R}}

\newcommand{\etal}{\emph{et al.\xspace}}

\newcommand{\query}{Q}
\newcommand{\objectSet}{\mathcal{O}}
\newcommand{\seeds}{\objectSet^*}
\newcommand{\bd}{\partial}
\newcommand{\vsegs}{\textsf{V}(\objectSet)}
\newcommand{\hsegs}{\textsf{H}(\objectSet)}
\newcommand{\vrects}{\textsf{V}(\query)}
\newcommand{\hrects}{\textsf{H}(\query)}
\newcommand{\ds}{\mathcal{D}}
\newcommand{\tree}{\mathcal{T}}
\newcommand{\node}{\nu}
\newcommand{\unionset}{\mathcal{U}}

\newcommand{\proj}[1]{\overline{#1}}

\begin{document}

\title{Finding Pairwise Intersections Inside a Query Range\thanks{M.~de Berg and A.~D.~Mehrabi were supported by the
       Netherlands Organization for Scientific Research (NWO) under grants 024.002.003 and 612.001.118, respectively.}}
\titlerunning{Finding Pairwise Intersections Inside a Query Range}

\author{Mark~de Berg\inst{1} \and Joachim Gudmundsson\inst{2} \and Ali~D.~Mehrabi\inst{1}}
\institute{Department of Computer Science, TU Eindhoven, the Netherlands
           \and Department of Computer Science, University of Sydney, Australia}

\maketitle

\begin{abstract}
  We study the following problem: preprocess a set $\objectSet$ of objects into
  a data structure that allows us to efficiently report all pairs of objects from $\objectSet$
  that intersect inside an axis-aligned query range~$\query$. We present data structures
  of size $O(n\polylog n)$ and with query time $O((k+1)\polylog n)$ time, where $k$
  is the number of reported pairs, for two classes of objects in the plane:
  axis-aligned rectangles and objects with small union complexity.
  For the 3-dimensional case where the objects and the query range
  are axis-aligned boxes in~$\Reals^3$, we present a data structures of size $O(n\sqrt{n}\polylog n)$ and
  query time $O((\sqrt{n}+k)\polylog n)$. When the objects
  and query are fat, we obtain $O((k+1)\polylog n)$ query time using $O(n\polylog n)$ storage.
\end{abstract}

\section{Introduction}
The study of geometric data structures is an important subarea within computational
geometry, and range queries form one of the most widely studied topics
within this area~\cite{ae-grsir-99,go-hdcg-04}.
In a range query, the goal is to report or count all points from a given set $\objectSet$
that lie inside a query range~$\query$. The more general version, where
$\objectSet$ contains other objects than just points and the goal is to
report all objects intersecting~$\query$, is often called intersection searching
and it has been studied extensively as well.

A common characteristic of the range-searching and intersection-searching problems
studied so far, is that whether an object $o_i\in\objectSet$ should be reported
(or counted) depends only on $o_i$ and~$\query$. In this paper we study a range-searching
variant where we are interested in reporting \emph{pairs} of objects that satisfy
a certain criterion. In particular, we want to preprocess a set $\objectSet=\{o_1,\ldots,o_n\}$
of $n$ objects in the plane such that, given a query range~$\query$, we can efficiently report all
pairs of objects $o_i,o_j$ that intersect inside~$\query$.  An obvious approach is
to precompute all intersections between the objects and store the intersections
in a suitable intersection-searching data structure. This may give fast query times,
but in the worst case any two objects intersect, so $\Omega(n^2)$ is a lower bound on
the storage for this approach. The main question is
thus: can we achieve fast query times with a data structure that uses subquadratic
(and preferably near-linear) storage in the worst case?

We answer this question affirmatively when
$\query$ is an axis-aligned rectangle in the plane and the objects are either
axis-aligned rectangles or objects with small union complexity.
For axis-aligned rectangles our data structure uses $O(n\log n)$ storage
and has $O((k+1)\log n\log^* n)$ query time,\footnote{Here
$\log^* n$ denotes the iterated logarithm.}
where $k$ is the number of reported pairs of objects.
Our data structure for classes of objects with small union complexity---disks and other types of fat
objects are examples---uses $O(U(n)\log n)$ storage,
where $U(n)$ is maximum union complexity of $n$ objects
from the given class, and it has $O((k+1)\log^2 n)$ query time.
We also consider a 3-dimensional extension of the planar case, where
the range $\query$ and the objects in $\objectSet$ are axis-aligned boxes.
Our data structures for this setting has size $O(n\sqrt{n}\log n)$ and query
time $O((k+1)\log^2\log^* n)$. For the special case where the query range
and the objects are fat, we present a data structure of $O(n\log^2 n)$ size
and $O((k+1)\log^2 n\log^* n)$ query time.


\section{Axis-aligned objects}\label{sec:segments+rectangles}
In this section we study the case where the set $\objectSet$ is a set of $n$
axis-aligned rectangles in the plane or boxes in~$\Reals^3$.
Our approach for these cases is the same and uses the following two-step query process.
\myenumerate{
\item \label{step:find-seeds} Compute a \emph{seed set} $\seeds(\query)\subseteq \objectSet$ of
      objects such that the following holds:
      for any two objects $o_i,o_j$ in $\objectSet$ such that $o_i$ and $o_j$
      intersect inside~$\query$, at least one of $o_i,o_j$ is in $\seeds(\query)$.
\item \label{step:report-answers} For each seed object $o_i\in\seeds(\query)$,
      perform an intersection query with the range $o_i\cap \query$
      in the set $\objectSet$, to find all objects $o_j\neq o_i$ intersecting $o_i$ inside~$\query$.
}
To make this approach efficient, we need that the seed set~$\seeds(\query)$ does not
contain too many objects that do not give an answer in Step~\ref{step:report-answers}.
For the planar case our seed set will satisfy $|\seeds(\query)| =O(1+k)$,
where $k$ denotes the number of pairs of objects in $\objectSet$ that
intersect inside~$\query$, while for the 3-dimensional case
we will have $|\seeds(\query)| =O(\sqrt{n}+k)$.

\subsection{The planar case}\label{subsec:planarCase}
\paragraph{Axis-aligned segments.}  
As a warm-up exercise we start with the case where
$\objectSet$ consists of axis-aligned segments.
Let $\objectSet =\{s_1,\ldots,s_n\}$ be a set of axis-aligned segments, and
let $\vsegs$ and $\hsegs$ denote the set of vertical and horizontal
segments in $\objectSet$, respectively. We assume for simplicity that we are
only interested in intersections between horizontal and vertical segments;
the solution can easily be adapted to the case where we also want to report
intersections between two horizontal (or two vertical) segments.
\medskip

The key to our approach is to be able to efficiently find the seed set $\seeds(\query)$.
To this end, during the preprocessing we compute an $O(n)$-sized subset $W$ of the intersection
points in $\objectSet$. We call intersection points in $W$ \emph{witnesses}. The witness set $W$
is defined as follows:
for each line segment $s_i\in\vsegs$ we put the topmost and bottommost
intersection points of $s_i$ with a segment from $\hsegs$ (if any) into $W$;
for each line segment $s_i\in\hsegs$ we put the leftmost and rightmost
intersection points of $s_i$ with a segment from $\vsegs$ (if any) into $W$.
Since we take at most two witness points for each line segment, the size of $W$ is clearly at most~$2n$.

Our data structure to find the seed set $\seeds(\query)$ now consists of three
components: First, we store~$W$ in a data structure $\ds_1$
for 2-dimensional orthogonal range reporting. Second, we
store $\vsegs$ in a data structure $\ds_2$ that allows us to decide
if there are any segments that completely cross the query rectangle $\query$ from
top to bottom, and that can report all such segments.
Third, we store $\hsegs$ in a data structure $\ds_3$ that allows us to decide
if there are any segments that completely cross the query rectangle $\query$ from
left to right.

Step~\ref{step:find-seeds} of the query procedure,
where we compute $\seeds(\query)$, proceeds as follows.
\myenumerate{
\item[1(i)] Perform a query in $\ds_1$ to find all witness points inside~$\query$.
      For each reported witness point, insert the corresponding segment into $\seeds(\query)$.
\item[1(ii)] Perform queries in $\ds_2$ and $\ds_3$ to decide if the number of segments crossing
      $\query$ completely from top to bottom, and the number of segments crossing
      $\query$ completely from left to right, are both non-zero. If so,
      report all segments crossing completely from top to bottom,
      and put them into~$\seeds(\query)$.
}
\begin{lemma}\label{lem:seg-query-correctness}
Let $s_i,s_j$ be two segments in $\objectSet$ such that $s_i\cap s_j \in \query$.
Then at least one of $s_i,s_j$ is put into $\seeds(\query)$ by the above query procedure.
\end{lemma}
\begin{myproof}
If $s_i$ crosses $\query$ completely from left to right and
$s_j$ crosses $\query$ completely from top to bottom (or vice versa),
then one of them will be put into $\seeds(\query)$ in Step~1(ii).
Otherwise at least one of the segments, say $s_i$, has an endpoint~$v$ inside~$\query$.
But then the intersection point on $s_i$ closest to~$v$, which is a witness point,
must lie inside $\query$. Hence, $s_i$ is put into $\seeds(\query)$ in Step~1(i).
\end{myproof}
In Step~\ref{step:report-answers} of the query procedure we need to
report, for each segment~$s_i$ in the seed set $\seeds(\query)$, the segments
$s_j\in \objectSet$ intersecting $s_i\cap\query$. Thus we store $\objectSet$
in a data structure $\ds_4$ that can report all segments intersecting
an axis-aligned query segment. Putting everything together we obtain the following theorem.
\begin{theorem}\label{thm:seg}
Let $\objectSet$ be a set of $n$ axis-aligned segments in the plane. Then there
is a data structure that uses $O(n\log n)$ storage and can report,
for any axis-aligned query rectangle~$\query$,
all pairs of segments $s_i,s_j$ in $\objectSet$ such that $s_i$ intersects $s_j$
inside $\query$
in $O((k+1)\log n\log^* n)$ time, where $k$ denotes the number of answers.
\end{theorem}
\begin{myproof}
For the data structure $\ds_1$ on the set~$W$ we can take a standard 2-dimensional
range tree~\cite{bcko-cgaa-08}, which uses $O(n\log n)$ storage.
If we apply fractional cascading~\cite{bcko-cgaa-08},
reporting the witness points inside $\query$ takes $O(\log n+\mbox{\#answers})$ time.
For $\ds_2$ (and, similarly, $\ds_3$) we note that a vertical segment $s_i := x_i \times [y_i,y'_i]$
crosses $\query := [x_{\query},x'_{\query}]\times [y_{\query},y'_{\query}]$
if and only if the point $(x_i,y_i,y'_i)$ lies in the range
$[x_{\query},x'_{\query}]\times [-\infty,y_{\query}]\times[y'_{\query},\infty]$.
Hence, we can use the data structure of Subramanian and Ramaswamy~\cite{subramanian95P-range},
which uses $O(n\log n)$ storage and has $O(\log n\log^* n + \mbox{\#answers})$ query time.
Hence, the supporting data structures for Step~\ref{step:find-seeds} use $O(n\log n)$ storage,
and finding the seed set takes $O(\log n \log^* n +|\seeds(\query)|)$ time.

It remains to analyze Step~\ref{step:report-answers} of the query procedure.
First notice that the problem of finding for a given $s_i \in \seeds(\query)$
all $s_j\in \objectSet$ such that $s_i\cap \query$ intersects $s_j$,
is the same range-searching problem as Step~1(ii), except that the query range
is a line segment this time.
Hence, we again transform the problem to a 3D range-searching problem on points and use
the data structure of Subramanian and Ramaswamy~\cite{subramanian95P-range}.
Thus the running time of Step~\ref{step:report-answers} is
$\sum_{s_i\in\seeds(\query)} O(\log\log^* n + k_i)$,
where $k_i$ denotes the number of segments in $\objectSet$ that intersect $s_i$ inside $\query$.
Since $|\seeds(\query)|\leq 2k$ where $k$ is the total number of reported
pairs---each segment in $\seeds(\query)$ intersects at least one other
segment inside $\query$ and for every reported pair we put at most two segments
into the seed set---the time for Step~\ref{step:report-answers} is
$O(\lvert \seeds(\query)\rvert\log n \log^* n + k) = O((k+1)\log n\log^* n)$.
\end{myproof}

\paragraph{Axis-aligned rectangles.}
We now extend our approach to axis-aligned rectangles.
Let $\objectSet =\{r_1,\ldots,r_n\}$ be a set of axis-aligned rectangles in the plane.
Similar to the case of axis-aligned segments we need to find the seed set $\seeds(\query)$ efficiently.
\begin{wrapfigure}{r}{37mm}
  \vspace*{-2mm}
  \includegraphics{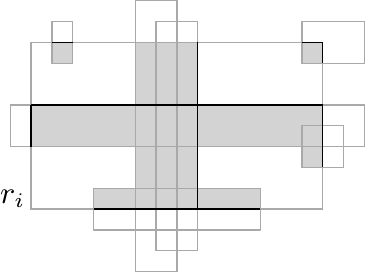}
  \captionof{figure}{Gray areas are intersections with~$r_i$,
                     black segments indicate witness segments.}
  \label{fig:test1}
  \vspace*{-6mm}
\end{wrapfigure}

As before, we first define a witness set~$W$. The witnesses in $W$ are
now axis-aligned segments rather than just points.
For each rectangle $r_i\in \objectSet$ we define at most ten witness segments, two
for each edge of $r_i$ and two in the interior of $r_i$, as follows---see
also Fig.~\ref{fig:test1}.
Let $e$ be an edge of $r_i$, and
consider the set~$S(e) := e \cap \left(\cup_{j\neq i} r_j\right)$, that is,
the part of $e$ covered by the other rectangles. The set $S(e)$ consists of a
number of sub-edges of~$e$.
If $e$ is vertical then we add the topmost and bottommost sub-edge from $S(e)$ (if any) to~$W$;
if $e$ is horizontal we add the leftmost and rightmost sub-edge to~$W$.
The two witness segments in the interior of $r_i$ are defined as follows.
Suppose there are vertical edges (belonging to other rectangles~$r_j$)
completely crossing $r_i$ from top to bottom. Then we put $e'\cap r_i$ into~$W$,
where $e'$ is the rightmost such crossing edge. Similarly, we put into $W$
the topmost horizontal edge $e''$ completely crossing $r_i$ from left to right.
Our data structure to find the seed set $\seeds(\query)$ now consists of the following components.
\myitemize{
\item We store the witness set $W$ in a data structure $\ds_1$ that allows us to report
      the set of segments that intersect the query rectangle $\query$.
\item We store the vertical edges of the rectangles in $\objectSet$ in a
      data structure~$\ds_2$ that allows us to decide if the set $\vrects$
      of edges that completely cross a query rectangle $\query$ from
      top to bottom, is non-empty.
      The data structure should also be able to report all (rectangles corresponding to)
      the edges in~$\vrects$.
\item We store the horizontal edges of the rectangles in $\objectSet$ in
      a data structure~$\ds_3$ that allows us to decide if the set $\hrects$
      of edges that completely cross a query rectangle $\query$ from
      left to right, is non-empty.
\item We store $\objectSet$ in a data structure $\ds_4$ that allows us to report
      the set of rectangles that contain a query point~$q$.
}
Step~\ref{step:find-seeds} of the query procedure,
where we compute $\seeds(\query)$, proceeds as follows.
\myenumerate{
\item[1(i)] Perform a query in $\ds_1$ to find all witness segments intersecting~$\query$.
      For each reported witness segment, insert the corresponding rectangle into $\seeds(\query)$.
\item[1(ii)] Perform queries in $\ds_2$ and $\ds_3$ to decide if the sets $\vrects$
      and $\hrects$ are both non-empty. If so, report all rectangles
      corresponding to edges in  $\vrects$ and put them into~$\seeds(\query)$.
\item[1(iii)] For each corner point $q$ of $\query$, perform a query in $\ds_4$ to report
      all rectangles in $\objectSet$ that contain $q$, and put them into~$\seeds(\query)$.
}
The next lemma can be proved using a case analysis---see the Appendix~\ref{sec:appendix-A}.
\begin{lemma}\label{lem:rec-query-correctness}
Let $r_i,r_j$ be two rectangles in $\objectSet$ such that
$(r_i\cap r_j) \cap \query \neq \emptyset$.
Then at least one of $r_i,r_j$ is put into $\seeds(\query)$ by the above query procedure.
\end{lemma}
%
In the second part of the query procedure we need to report,
for each rectangle~$r_i$ in the seed set~$\seeds(\query)$, the rectangles
$r_j\in \objectSet$ intersecting $r_i\cap\query$. Thus we store
$\objectSet$ in a data structure $\ds_5$ that can report all rectangles intersecting
a query rectangle. Putting everything together we obtain the following theorem.
\begin{theorem}\label{thm:rect}
Let $\objectSet$ be a set of $n$ axis-aligned rectangles in the plane. There
is a data structure that uses $O(n\log n)$ storage and can report,
for any axis-aligned query rectangle~$\query$,
all pairs of rectangles $r_i,r_j$ in $\objectSet$ such that $r_i$ intersects $r_j$
inside $\query$
in $O((k+1)\log n\log^* n)$ time, where $k$ denotes the number of answers.
\end{theorem}
\begin{myproof}
For the data structure $\ds_1$ on the set~$W$ we use the data structure developed by Edelsbrunner~\etal~\cite{edelsbrunner84linesegmentswindowing},
which uses $O(n\log n)$ preprocessing time and storage, and has $O(\log n + \mbox{\#answers})$ query time.

Data structure $\ds_2$ (and, similarly, $\ds_3$) answers the same type of query we needed when $\objectSet$ contains segments.
Hence, we can use the same data structure~\cite{subramanian95P-range} which uses $O(n\log n)$ space and has
$O(\log n\log^ *n + \mbox{\#answers})$ query time. For data structure $\ds_4$ we use the point-enclosure data structure developed
by Chazelle~\cite{Chazelle86filteringsearch}, which uses $O(n)$ storage and can be used to report all rectangles in $\objectSet$
containing a query point in $O(\log n + \mbox{\#answers})$ time.

The analysis of Step~\ref{step:report-answers} is similar to the analysis
for the case of axis-aligned segments,
except that we now have $|\seeds(\query)|\leq 2k+4$, where $k$ is the total number
of pairs of rectangles that will be reported; the extra term ``+4'' is because
in Step~\ref{step:find-seeds}(iii) we may report at most one
rectangle per corner of~$\query$ that does not have an intersection inside~$\query$.
Again, finding the rectangles in $\objectSet$ intersecting $r_i\cap \query$,
for a given $r_i\in \seeds(\query)$, can be done in $O(\log n\log^ *n + \mbox{\#answers})$,
leading to an overall query time of $O((k+1)\log n\log^* n)$.
\end{myproof}

\subsection{The 3-dimensional case}
We now study the case where the set $\objectSet$ of objects and the
query range $\query$ are axis-aligned boxes in $\Reals^3$. We first present a solution
for the general case, and then an improved solution for the special case where the input
as well as the query are cubes. Both solutions use the same query strategy as above:
we first find a seed set $\seeds(\query)$ that contains at least one object $o_i$ from
every pair that intersects inside~$\query$ and then we find all other objects
intersecting~$o_i$ inside $\query$.

\paragraph{The general case.}\label{para:3d-general}
Let $\objectSet := \{b_1,\ldots,b_n\}$ be a set of axis-aligned boxes.
The pairs of boxes $b_i,b_j$ intersecting inside $\query$ come in three types:
(i)~$b_i\cap b_j$ fully contains $\query$,
(ii)~$b_i\cap b_j$ lies completely inside $\query$,
(iii)~$b_i\cap b_j$ intersects a face of $\query$.

Type~(i) is easy to handle without using seeds sets:
we simply store $\objectSet$ in a data structure for 3-dimensional
point-enclosure queries~\cite{Chazelle86filteringsearch},
which allows us to report all boxes $b_i\in\objectSet$ containing
a query point in $O(\log^2 n+\mbox{\#answers})$ time. If we query this structure
with a corner~$q$ of $\query$ and report all pairs of boxes containing~$q$
then we have found all intersecting pairs of Type~(i).
\begin{lemma}\label{le:intersection-contains-Q}
We can find all intersecting pairs of boxes of Type~(i) in $O(\log^2 n+k)$
time, where $k$ is the number of such pairs, with a structure
of size~$O(n\log n)$.
\end{lemma}
For Type~(ii) we proceed as follows. Note that a vertex of $b_i\cap b_j$
is either a vertex of $b_i$ or $b_j$, or it is the intersection of an edge~$e$
of one of these two boxes and a face~$f$ of the other box. To handle the first case
we create a set $W$ of witness points, which contains
for each box $b_i$ all its vertices that are contained in at least one other box.
We store~$W$ in a data structure for 3-dimensional orthogonal range reporting~\cite{subramanian95P-range}.
In the query phase we then query this data structure with~$\query$, and put
all boxes corresponding to the witness vertices inside~$\query$ into the seed
set~$\seeds(\query)$.
For the second case we show next how to find the intersecting pairs $e,f$
where $e$ is a vertical edge (that is, parallel to the $z$-axis)
and $f$ is a horizontal face (that is, parallel to the $xy$-plane);
the intersecting pairs with other orientations can be found
in a similar way.

Let $E$ be the set of vertical edges of the boxes
in $\objectSet$ and let $F$ be the set of horizontal faces.
We sort $F$ by $z$-coordinate---we assume for
simplicity that all $z$-coordinates of the faces are distinct---and
partition $F$ into $O(\sqrt{n})$ \emph{clusters}: the cluster~$F_1$
contains the first $\sqrt{n}$ faces in the sorted order,
the second cluster~$F_2$ contains the next $\sqrt{n}$ faces, and so on.
We call the range between the minimum and maximum $z$-coordinate
in a cluster its \emph{$z$-range}.
For each cluster $F_i$ we store, besides its $z$-range and the set $F_i$ itself,
the following information. Let $E_i\subseteq E$ be the subset of edges
that intersect at least one face in~$F_i$, and let $\proj{E_i}$ denote
the set of points obtained by projecting the edges in $E_i$ onto the $xy$-plane.
We store $\proj{E_i}$ in a data structure $\ds(\proj{E_i})$ for 2-dimensional orthogonal range
reporting. Note that an edge $e\in E$ intersects at least one face $f\in F_i$ inside $\query$
if and only if $e\in E_i$ and $\proj{e}$ lies in $\proj{\query}$,
the projection of $\query$ onto the $xy$-plane.

A query with a box~$\query=[x_1:x_2]\times [y_1:y_2]\times[z_1:z_2]$
is now answered as follows. We first find the clusters $F_i$ and $F_j$ whose
$z$-range contains $z_1$ and~$z_2$, respectively, and we put (the boxes
corresponding to) the faces in these clusters into the seed set $\seeds(\query)$.
Next we perform, for each $i<t<j$, a query with the projected range
$\proj{\query}$ in the data structure~$\ds(\proj{E_i})$.
For each of the reported points $\proj{e}$ we put the box corresponding
to the edge $e$ into the seed set~$\seeds(\query)$. Finally, we remove
any duplicates from the seed set.

We obtain the following lemma, whose proof is in the Appendix~\ref{sec:appendix-A}.
\begin{lemma}\label{le:typeii}
Using a data structure of size $O(n\sqrt{n}\log n)$ we can find in time
$O(\log n\log^* n + k)$ a seed set $\seeds(\query)$ of $O(\sqrt{n}+k)$ boxes
containing at least one box from every intersecting pair of Type~(ii),
where $k$ is the number of such pairs.
\end{lemma}
%
It remains to handle the Type~(iii) pairs, in which
$b_i\cap b_j$ intersects a face of~$\query$.
We describe how to find the pairs such that $b_i\cap b_j$ intersects
the bottom face of $\query$; the pairs
intersecting the other faces can be found in a similar way.

We first sort the $z$-coordinates of the horizontal
faces of the boxes in $\objectSet$. For $1\leq i\leq 2\sqrt{n}$,
let $h_i$ be a horizontal plane containing the $i\sqrt{n}$-th horizontal face
in the ordering. These planes partition $\Reals^3$ into $O(\sqrt{n})$ horizontal
slabs $\Sigma_0,\ldots,\Sigma_{2\sqrt{n}+1}$.
We call a box $b\in \objectSet$ \emph{short} for a slab $\Sigma_i$ if
it has a horizontal face inside~$\Sigma_i$, and we call it \emph{long}
if it completely crosses~$\Sigma_i$. For each $\Sigma_i$, we store the short
boxes in a list. We store the projections of the long boxes onto
the $xy$-plane in a data structure $\ds(\Sigma_i)$ for
the 2-dimensional version of the problem, namely
the structure  Theorem~\ref{thm:rect}.

A query with the bottom face of~$\query$
is now answered as follows. We first find the slab $\Sigma_i$ containing the face.
We put all short boxes of $\Sigma_i$ into our seed set $\seeds(\query)$.
We then perform a query with $\proj{\query}$,
the projection of $\query$ onto the $xy$-plane, in the data structure $\ds(\Sigma_i)$.
For each answer we get from this 2-dimensional query---that is,
each pair of projections intersecting inside $\proj{Q}$---we directly
report the corresponding pair of long boxes. (There is no need to
go through the seed set for these pairs.)
This leads to the following lemma for the Type~(iii) pairs.
\begin{lemma}\label{le:typeii}
Using a data structure of size $O(n\sqrt{n}\log n)$ we can find
in time $O(\sqrt{n}+(k+1)\log^* n\log n)$ a seed set $\seeds(\query)$ of
$O(\sqrt{n})$ boxes plus a collection $B(\query)$ of pairs of boxes
intersecting inside $\query$
such that, for each pair of Type~(iii) boxes, either
at least one of these boxes is in $\seeds(\query)$ or $b_i,b_j$ is a pair in $B(\query)$.
\end{lemma}
In the second step of our query procedure we need to be able to report all boxes
$b_j\in\objectSet$ intersecting a query box $B$ of the form $\query \cap b_i$,
where $b_i\in \seeds(\query)$. Note that $B$ and $b_j$ intersect if and only if
their projections onto the $z$-axis intersect and their projections onto the $xy$-plane
intersect. Hence, we can answer the queries with a data structure $\ds^*$ whose main tree is a
(hereditary) segment tree~\cite{cegs94hereditary} and whose associated structures are the data
structure of Subramanian and Ramaswamy~\cite{subramanian95P-range}.
This leads to a structure using $O(n\log^2 n)$ storage and $O(\log^2 n \log^* n + \mbox{\#answers})$
query time.

Putting everything together we obtain the following theorem.
\begin{theorem}\label{thm:3d-general}
Let $\objectSet$ be a set of $n$ axis-aligned boxes in~$\Reals^3$. Then there
is a data structure that uses $O(n\sqrt{n}\log n)$ storage and that allows us to report,
for any axis-aligned query box~$\query$,
all pairs of boxes $b_i,b_j$ in $\objectSet$ such that $b_i$ intersects $b_j$
inside $\query$ in $O(\sqrt{n}+(k+1)\log^2 n\log^* n)$ time, where $k$ denotes the number of answers.
\end{theorem}

\paragraph{Fat boxes.}\label{para:3d-cube}
Next we obtain better bounds when the boxes in $\objectSet$ and the
query box~$\query$ are fat, that is, when their \emph{aspect ratio}---the ratio between
the length of the longest edge and the length of the shortest edge---is bounded by a constant~$\alpha$.
First we consider the case of cubes.

Let $\objectSet := \{c_1, \cdots, c_n\}$ be a set of $n$ cubes in $\Reals^3$ and
let $\query$ be the query cube.
We compute a set $W$ of witness points for each cube~$c_i$, as follows.
Let $e$ be an edge of~$c_i$, and consider the
set~$S(e) := e \cap \left(\cup_{j\neq i} c_j\right)$, that is,
the part of $e$ covered by the other cubes. We put the two extreme points from $S(e)$---in other words,
the two points closest to the endpoints of $e$---into~$W$.
Similarly, we assign each face $f$ of~$c_i$ at most four witness points, namely
points from~$S(f) := f\cap (\cup_{j\neq i} c_j)$ that are extreme in the directions parallel to~$f$.
For example, if $f$ is parallel to the $xy$-plane, then we take points of maximum
and minimum $x$-coordinate in $S(f)$ and points of maximum and minimum $y$-coordinate in $S(f)$ as witnesses.
%
We store~$W$ in a data structure $\ds_1$ for orthogonal range queries,
and we store $\objectSet$ in a data structure $\ds_2$ for point-enclosure queries.

\begin{figure}[t]
\centering
\begin{minipage}{0.46\textwidth}
  \centering
  \includegraphics{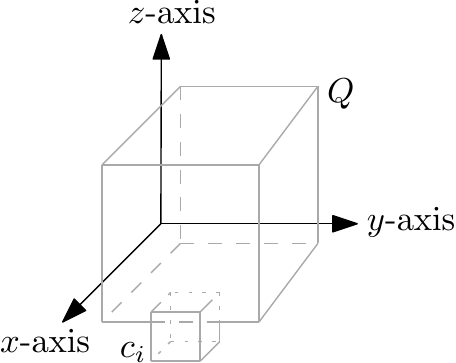}
  \captionof{figure}{Case~B in the proof of Lemma~\ref{lem:cube-query-correctness};
                     $c_j$ is not shown.}
  \label{fig:cube-query}
\end{minipage}
\qquad
\begin{minipage}{.46\textwidth}
  \centering
  \includegraphics{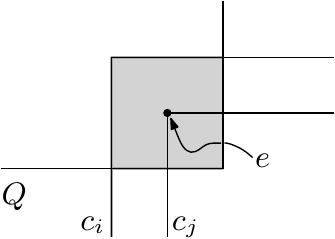}
  \captionof{figure}{Cross-section of~$\query$, $c_i$, and $c_j$ with a plane parallel to the $xz$-plane.
                     The gray area indicates $\query\cap c_i$ in the cross-section.}
  \label{fig:cube-query2}
\end{minipage}
\end{figure}

To compute $\seeds(\query)$ in the first phase of the query procedure,
we query $\ds_1$ to find all witness points inside~$\query$ and for
each reported witness point, we insert the corresponding cube into $\seeds(\query)$.
Furthermore, for each corner point $q$ of $\query$, we query $\ds_2$ to find
the cubes in $\objectSet$ that contain $q$, and we put them into~$\seeds(\query)$.
%
%
\begin{lemma}\label{lem:cube-query-correctness}
Let $c_i,c_j$ be two cubes in $\objectSet$ such that $(c_i\cap c_j) \cap \query \neq \emptyset$.
Then at least one of $c_i,c_j$ is put into $\seeds(\query)$ by the above query procedure.
\end{lemma}
\begin{myproof}
Suppose $c_i\cap c_j$ intersects $\query$, and assume without loss of generality that $c_i$ is
not larger than $c_j$. If $c_i$ or $c_j$ contains a corner $q$ of $\query$
then the corresponding cube will be put into the seed set when we perform a point-enclosure query with~$q$,
so assume $c_i$ and $c_j$ do not contain a corner. We have two cases.

\textsc{Case A:} $c_i$ does not intersect any edge of $\query$. Because $c_i$ and $\query$ are cubes, this
implies that $c_i$ is contained in~$\query$ or $c_i$ intersects exactly one face of $\query$. Assume
that $c_i$ intersects the bottom face of $\query$; the cases where $c_i$ intersects another face
and where $c_i$ is contained in $\query$ can be handled similarly.
We claim that at least one of the vertical faces of $c_i$ contributes a witness
point inside $\query$. To see this, observe that $c_j$ will intersect at least one vertical face, $f$, of $c_i$
inside $\query$, since $c_j$ intersects $c_i$ inside~$\query$ and $c_i$ is not larger than~$c_j$.
Hence, the witness point on $f$ with maximum $z$-coordinate will be inside $\query$.
Thus $c_i$ will be put into $\seeds(\query)$.

\textsc{Case B:} $c_i$ intersects one edge of $\query$. (If $c_i$ intersects more than one edge of $\query$
then it would contain a corner of~$\query$.) Assume without loss of generality that $c_i$ intersects the
bottom edge of the front face of $\query$; see Fig.~\ref{fig:cube-query}.
Observe that if $c_j$ intersects the top face of $c_i$ then the witness point of the face with minimum
$x$-coordinate is inside~$\query$. Similarly, if $c_j$ intersects the back face of~$c_i$
(the face parallel to the $yz$-plane and with minimum $x$-coordinate)
then the witness point of the face with maximum $z$-coordinate is inside~$\query$.
Otherwise, as illustrated in Fig~\ref{fig:cube-query2}, $c_j$ must have an edge~$e$
parallel to the $y$-axis that intersects $c_i$ inside $\query$, and one of the witness points on~$e$
will be inside $\query$---note that $e$ lies fully inside $\query$ because $c_j$ does not
contain a corner of~$\query$.
\end{myproof}

To adapt the above solution to boxes of aspect ratio at most $\alpha$, we cover each
box $b_i\in\objectSet$ by $O(\alpha^2)$ cubes, and preprocess the resulting collection $\widetilde{\objectSet}$
of cubes as described above, making sure we do not introduce witness
points for pairs of cubes used in the covering of the same box~$b_i$.
To perform a query, we cover~$\query$ by $O(\alpha^2)$ query cubes and compute
a seed set for each query cube. We take the union of these seed sets, replace the cubes
from $\widetilde{\objectSet}$ in the seed set by the corresponding boxes in $\objectSet$, and
filter out duplicates. This gives us our seed set $\seeds(\query)$ for the second phase of the query procedure.

In the second phase we take each $b_i\in\seeds(\query)$
and report all $b_j\in \objectSet$ intersecting~$b_i\cap\query$, using
the data structure $\ds^*$ described in Subsection~\ref{para:3d-general}.
We obtain the following theorem.
\begin{theorem}\label{thm:3d-cube}
Let $\objectSet$ be a set of $n$ axis-aligned boxes in~$\Reals^3$ of aspect ratio at most~$\alpha$.
Then there is a data structure that uses $O(\alpha^2 n\log^2 n)$ storage and that allows us to report,
for any axis-aligned query box~$\query$ of aspect ratio at most $\alpha$,
all pairs of cubes $c_i,c_j$ in $\objectSet$ such that $c_i$ intersects $c_j$
inside $\query$ in $O(\alpha^2(k+1)\log^2\log^* n)$ time, where $k$ denotes the number of answers.
\end{theorem}
\begin{myproof}
The data structures $\ds_1$ and $\ds_2$ can be implemented such that they
use $O(n\log n)$ storage, and have $O(\log n\log^* n + \mbox{\#answers})$
and $O(\log^2 n + \mbox{\#answers})$ query time, respectively~\cite{subramanian95P-range,Chazelle86filteringsearch}.
In Step~\ref{step:report-answers} of the query procedure
we use the data structure $\ds^*$ of Subsection~\ref{para:3d-general}, which uses $O(n\log^2 n)$ storage and has
$O(\log^2\log^* n + \mbox{\#answers})$ query time.
The conversion of boxes of aspect ratio $\alpha$ to cubes give an additional factor $O(\alpha^2)$.
%
\end{myproof}

\section{Objects with small union complexity in the plane}\label{sec:const-complexity-objs}
In the previous section we presented efficient solutions for the case where
$\objectSet$ consists of axis-aligned rectangles. In this section
we obtain results for classes of constant-complexity objects (which may
have curved boundaries) with small union
complexity. More precisely, we need that  $U(n)$,
the maximum union complexity of any set of $n$ objects from the class, is small.
This is for instance the case for disks (where $U(m)=O(m)$~\cite{klps-oujrctmapo-86}) and
for locally fat objects (where $U(m)=m2^{O(\log^* m)}$~\cite{abes-ibulfo-14}).

In Step~\ref{step:report-answers} of the query algorithm of the
previous section, we performed a range query with $o_i\cap \query$
for each~$o_i\in\seeds(\query)$. When we are dealing with arbitrary objects,
this will be expensive, so we modify our query procedure.
\myenumerate{
\item \label{step:find-seeds-arbitrary} Compute a seed set
      $\seeds(\query)\subseteq \objectSet$ of objects such that,
      for any two objects $o_i,o_j$ in $\objectSet$
      intersecting inside~$\query$, both $o_i$ and $o_j$ are in $\seeds(\query)$.
\item \label{step:report-answers-arbitrary} Compute all intersecting pairs of objects
      in the set $\{ o_i\cap \query : o_i \in \seeds(\query) \}$ by a plane-sweep
      algorithm.
}
%
%
Next we describe how to efficiently find~$\seeds(\query)$, which
should contain all objects intersecting at least one other object inside $\query$,
when the union complexity $U(n)$ is small.
For each object $o_i\in\objectSet$ we define
$
o_i^* := \bigcup_{{o_j\in\objectSet},{j \neq i}} (o_i \cap o_j)
$
as the union of all intersections between $o_i$ and all other objects in $\objectSet$.
Let $|o_i^*|$ denote the complexity (that is, number of vertices and edges) of $o_i^*$.
\begin{lemma}\label{lem:leavesComplexities}
$\sum_{i=1}^n |o_i^*| = O(U(n))$.
\end{lemma}
\begin{myproof}
Consider the arrangement induced by the objects in~$\objectSet$.
We define the \emph{level} of a vertex $v$ in this arrangement as the
number of objects from $\objectSet$ that contain~$v$ in their interior.
We claim that every vertex of any $o_i^*$ is a level-0 or level-1 vertex.
Indeed, a level-$k$ vertex for $k>1$ is in interior of more than one
object, which is easily seen to imply
that it cannot be a vertex of any~$o_i^*$.

Since the level-0 vertices are exactly the vertices of the union of $\objectSet$,
the total number of level-0 vertices is $U(n)$. It follows
from the Clarkson-Shor technique~\cite{cs-aprscg-89} that the number of level-1 vertices
is $O(U(n))$ as well. The lemma now follows, because
each level-0 or level-1 vertex contributes to at most two different $o_i^*$'s.
\end{myproof}
Our goal in Step~\ref{step:find-seeds-arbitrary} is to find all objects~$o_i$ such that
$o_i^*$ intersects~$\query$. To this end consider the connected components of~$o_i^*$.
If $o_i^*$ intersects~$\query$ then one of these components lies
completely inside~$\query$ or an edge of $\query$ intersects~$o_i^*$.
\begin{lemma}
We can find all $o_i^*$ that have a component completely inside $\query$ in $O(\log n + k)$
time, where $k$ is the number of pairs of objects that intersect inside~$\query$,
with a data structure that uses $O(U(n)\log n)$ storage.
\end{lemma}
\begin{myproof}
For each $o_i$, take an arbitrary representative point inside each component of~$o_i^*$,
and store all the representative points in a structure for orthogonal range reporting.
By Lemma~\ref{lem:leavesComplexities} we store $O(U(n))$ points, and so
the structure for orthogonal range reporting uses $O(U(n)\log n)$ storage.

The query time is $O(\log n + t)$, where $t$ is the number of representative points
inside~$\query$. This implies the query time is $O(\log n + k)$,
because if $o_i^*$ has $t_i$ representative
points inside~$\query$ then $o_i$ intersects $\Omega(t_i)$ other objects inside~$\query$.
This is true because the objects have constant complexity, so a single object
$o_j$ cannot generate more than a constant number of components of~$o_i^*$.
\end{myproof}
Next we describe a data structure for reporting all $o_i^*$
intersecting a vertical edge of~$\query$; the horizontal
edges of~$\query$ can be handled similarly.
The data structure is a balanced binary tree $\tree$,
whose leaves are in one-to-one correspondence to the
objects in $\objectSet$. For an (internal or leaf) node $\node$ in $\tree$,
let $\tree(\node)$ denote the subtree rooted at~$\node$ and let
$\objectSet(\node)$ denote the set of objects corresponding to the leaves of $\tree(\node)$.
Define $\unionset(\node) := \cup_{o_i \in \objectSet(\node)} o_i^*$.
At node $\node$, we store a point-location data structure~\cite{egs86pointlocation} on the trapezoidal
map of $\unionset(\node)$. (If the objects are curved, then the
``trapezoids'' may have curved top and bottom edges.)
\begin{lemma}
The tree $\tree$ uses $O(U(n)\log n)$ storage and allows us to
report all $o_i^*$ intersecting a vertical edge $s$ of $\query$ in $O((t+1)\log^2 n)$
time, where $t$ is the number of answers.
\end{lemma}
\begin{myproof}
To report all $o_i^*$ intersecting~$s$ we walk down $\tree$,
only visiting the nodes~$\node$ such that $s$ intersects~$\unionset(\node)$.
This way we end up in the leaves corresponding to the~$o_i^*$
intersecting~$s$. To decide if we have to visit a child~$\node$
of an already visited node, we do a point location with both endpoints
of $s$ in the trapezoidal map of~$\unionset(\node)$. Now $s$ intersects
$\unionset(\node)$ if and only if one of these endpoints lies in a trapezoid
inside $\unionset(\node)$ and/or the two endpoints lie in different
trapezoids. Thus we spend $O(\log n)$ time for the decision.
Since we visit $O(k\log n)$ nodes, the total query time is as claimed.

To analyze the storage we claim that the sum of
the complexities of $\unionset(\node)$ over all nodes $\node$
at any fixed height of $\tree$ is $O(U(n))$. The bound on the storage then
follows because the point-location data structures
take linear space~\cite{egs86pointlocation} and the height of~$\tree$
is $O(\log n)$. It remains to prove the claim.
Consider a node $\node$ at a given height~$h$ in $\tree$.
Lemma~\ref{lem:complexity-U(v)} in Appendix~\ref{sec:appendix-A} proves that each vertex in~$\unionset(\node)$ is either
a level-$0$ or level-$1$ vertex of the arrangement induced by the objects in $\objectSet(\node)$,
or a vertex of $o^*_i$, for some $o_i$ in $\objectSet(\node)$.
The proof of the claim then follows from the following two facts. First, the number of vertices of the former type is
$O(U(\lvert \objectSet(\node)\rvert))$, which sums to $O(U(n))$ over all nodes at height~$h$.
Second, by Lemma~\ref{lem:leavesComplexities} the number of vertices of the latter type over all nodes at height~$h$
sums to $O(U(n))$.
\end{myproof}
%
%
\begin{theorem}\label{thm:constantComplexityObjs}
Let $\objectSet$ be a set of $n$ constant-complexity objects in the plane
from a class of objects such that the maximum union complexity of any $m$ objects
from the class is $U(m)$.
Then there is a data structure that uses $O(U(n)\log n)$ storage and that allows us to report
for any axis-aligned query rectangle~$\query$, in $O((k+1)\log^2 n)$ time all pairs of objects
$o_i,o_j$ in $\objectSet$ such that $o_i$ intersects $o_j$ inside $\query$,
where $k$ denotes the number of answers.
\end{theorem}

\section{Concluding remarks}
We presented data structures for finding intersecting pairs of objects inside a query rectangle.
An obvious open problem is whether our bounds can be improved. In particular, one would hope that
better solutions are possible for 3-dimensional boxes, where we obtained $O((k+\sqrt{n})\polylog n)$
query time with $O(n\sqrt{n}\log n)$ storage. (It is possible to reduce the query time in our
solution to $O((k+m)\polylog n)$, for any $1\leq m\leq\sqrt{n}$, but at the cost of increasing the
storage to $O((n^2/m)\polylog n)$.)

Two settings where we have not been able to obtain efficient solutions are when
$\objectSet$ is a set of balls in $\Reals^3$, and when $\objectSet$ is a set
of arbitrary segments in the plane. Especially the latter setting seems challenging.
Indeed, consider the special case where $\objectSet$ consist of $n/2$ horizontal lines
and $n/2$ lines of slope~1. Suppose furthermore that the query is a vertical line~$\ell$ and that
we only want to check if  $\ell$ contains at least one intersection. A data structure for
this setting could be used to solve the following \textsc{3Sum}-hard problem: given three
sets of parallel lines, decide if there is a triple intersection~\cite{3sum}.
Thus it is unlikely that we can obtain a solution with (significantly) sublinear
query time and (significantly) subquadratic preprocessing time in the setting just described.
However, storage is not the same as preprocessing time. This raises the following question: is it
possible to obtain sublinear query time with subquadratic storage?

\newpage
\appendix
\section{Omitted proofs}\label{sec:appendix-A}

\setcounter{lemma}{1} 
\begin{lemma}
Let $r_i,r_j$ be two rectangles in $\objectSet$ such that
$(r_i\cap r_j) \cap \query \neq \emptyset$.
Then at least one of $r_i,r_j$ is put into $\seeds(\query)$ by the above query procedure.
\end{lemma}
\begin{myproof}
Let $I:=(r_i\cap r_j) \cap \query$. Each edge of $I$ is either contributed by
$r_i$ or $r_j$, or by~$\query$. Let $E(I)$ denote
the set of edges of $r_i$ and $r_j$ that contribute an edge to~$I$.
We distinguish two cases, with various subcases.
\medskip

\textsc{Case A:} At least one edge $e\in E(I)$ has an endpoint, $v$, inside $\query$.
Now the witness sub-edge on $e$ closest to $v$ must intersect $Q$ and,
hence, the corresponding rectangle will be put into
$\seeds(\query)$ in Step~\ref{step:find-seeds}(i).

\textsc{Case B:} All edges in $E(I)$ cross $\query$ completely.
We now have several subcases.

\textsc{Case B-1:} $\lvert E(I)\rvert \leq 1$.
Now $\query$ contributes at least three edges to $I$, so
at least one corner of~$I$ is a corner of~$\query$. Hence, both $r_i$ and $r_j$
are put into $\seeds(\query)$ in Step~\ref{step:find-seeds}(iii).

\textsc{Case B-2:} $\lvert E(I)\rvert \geq 3$.
Since each edge of $E(I)$ crosses $\query$ completely
and $\lvert E(I)\rvert \geq 3$, both $\vrects$ and $\hrects$
are non-empty. Thus at least one of $r_i$ and $r_j$
is put into $\seeds(\query)$ in Step~\ref{step:find-seeds}(ii).

\textsc{Case B-3:} $\lvert E(I)\rvert =2$.
Let $e_1$ and $e_2$ denote the segments in~$E(I)$. If one of
$e_1,e_2$ is vertical and the other is horizontal, we can
use the argument from Case~B-2. It remains to handle the
case where $e_1$ and $e_2$ have the same orientation, say vertical.

\textsc{Case B-3-i:} Edges $e_1$ and $e_2$ belong to the same rectangle, say $r_i$,
as in Fig.~\ref{fig:fig}.
\begin{figure}[t]
  \centering
  \includegraphics{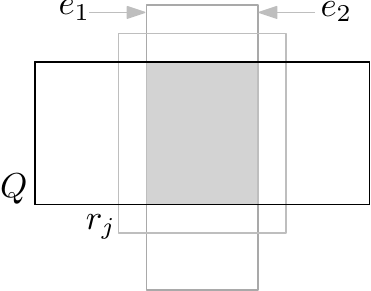}
  \captionof{figure}{A possible situation in Case~B-3-I.}
  \label{fig:fig}
\end{figure}
If $e_1$ has an endpoint, $v$, inside $r_j$,
then $e_1$ has a witness sub-edge starting at $v$ that intersects
$\query$, so $r_i$ is put into
$\seeds(\query)$ in Step~\ref{step:find-seeds}(i). If $r_j$ contains
a corner of $\query$ then $r_j$ will be put into
$\seeds(\query)$ in Step~\ref{step:find-seeds}(iii). In the remaining
case the right edge of $r_j$ crosses
$\query$ and there are vertical edges completely crossing $r_j$
(namely $e_1$ and $e_2$). Hence, the rightmost edge completely crossing
$r_j$, which is a witness for $r_j$, intersects~$\query$. Thus $r_j$
is put into $\seeds(\query)$ in Step~\ref{step:find-seeds}(i).

\textsc{Case B-3-ii:} Edge $e_1$ is an edge of $r_i$ and $e_2$ is an edge of~$r_j$ (or vice versa).
Assume without loss of generality that the $y$-coordinate of the top endpoint of $e_1$
is less than or equal to the $y$-coordinate of the top endpoint of $e_2$. Then
the top endpoint, $v$, of $e_1$ must lie in $r_j$, and so
$e_1$ has a witness sub-edge starting at $v$ that intersects~$\query$.
Hence, $r_i$ is put into $\seeds(\query)$ in Step~\ref{step:find-seeds}(i).
\end{myproof}

\setcounter{lemma}{3}
\begin{lemma}
Using a data structure of size $O(n\sqrt{n}\log n)$ we can find in time
$O(\log n\log^* n + k)$ a seed set $\seeds(\query)$ of $O(\sqrt{n}+k)$ boxes
containing at least one box from every intersecting pair of Type~(ii),
where $k$ is the number of such pairs.
\end{lemma}
\begin{myproof}
The Type~(ii) intersections $b_i \cap b_j$ either
have a vertex that is a vertex of $b_i$ or $b_j$ inside~$\query$, or they have
an edge-face pair intersecting inside~$\query$.
To find seed objects for the former pairs we used $O(n\log n)$ storage
and $O(\log n\log^* n + \mbox{\#answers})$ query time, and we put $O(k)$
boxes into the seed set.
For the latter pairs, we used an approach based
on clusters. For each cluster~$F_i$ we have a data structure $\ds(\proj{E_i})$
that uses $O(n\log n)$ storage, giving $O(n\sqrt{n}\log n)$ storage in total.
Besides the $O(\sqrt{n})$ boxes in the two clusters $F_i$ and $F_j$,
we put boxes into the seed set for the clusters $F_t$ with $i<t<j$,
namely when querying the data structures~$\ds(\proj{E_i})$.
This means that the same box may be put into $\seeds(\query)$ up to
$\sqrt{n}$ times. (Note that these duplicates are later removed.)
However, each copy we put into the seed set corresponds to a different
intersecting pair. Together with the fact that the query time in each
$\ds(\proj{E_t})$ is $O(\log n\log^* n+\mbox{\#answers})$ this means the
total query time and size of the seed set are as claimed.
\end{myproof}

\begin{figure}[t]
        \centering
        \begin{subfigure}[b]{0.28\textwidth}
                \includegraphics{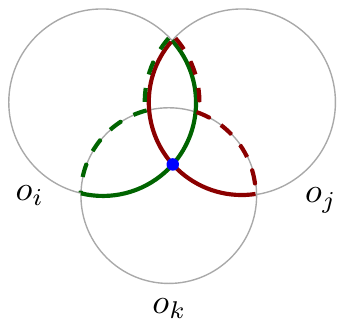}
                \caption{Case A in the proof of Lemma~\ref{lem:complexity-U(v)}.}
                \label{fig:regular-irregularA}
        \end{subfigure}%
        \hspace{2cm}
        ~ 
        \begin{subfigure}[b]{0.28\textwidth}
                \includegraphics{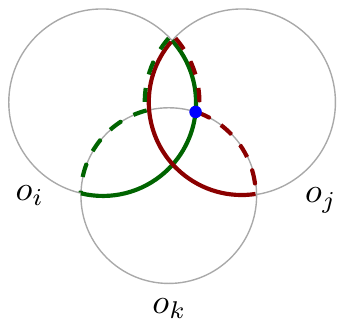}
                \caption{Case B in the proof of Lemma~\ref{lem:complexity-U(v)}.}
                \label{fig:regular-irregularB}
        \end{subfigure}
        \caption{Different cases in the proof of Lemma~\ref{lem:complexity-U(v)}.
                 To simplify the presentation we assumed the objects are disks.
                 $o^*_i$ and $o^*_j$ are surrounded by dark green
                 and dark red, respectively. Regular arcs are in solid and irregular arcs are in dashed.
                 The blue vertex refers to vertex $u$ in the proof.}
        \label{fig:regular-irregular}
\end{figure}

\begin{lemma}\label{lem:complexity-U(v)}
Each vertex in~$\unionset(\node)$ is either a level-$0$ or level-$1$ vertex of the arrangement induced
by the objects in $\objectSet(\node)$, or a vertex of $o^*_i$, for some $o_i$ in $\objectSet(\node)$.
\end{lemma}

\begin{myproof}
Define $\objectSet^*(\node) := \{ o_i^* : o_i \in \objectSet(\node) \}$.
Any vertex $u$ of $\unionset(\node)$ that is not a vertex of some $o_i^* \in \objectSet^*(\node)$
must be an intersection of the boundaries of some  $o_i^*,o_j^*\in \objectSet(\node)$.
Note that the boundary $\bd o_i^*$ of an object $o_i^*$ consists of two types of pieces:
\emph{regular arcs}, which are parts of the boundary of $o_i$ itself, and \emph{irregular arcs},
which are parts of the boundary of some other object $o_k$. To bound the number of vertices
of $\unionset(\node)$ of the form $\bd o_i^* \cap \bd o_j^*$ we now distinguish three cases.

\textsc{Case A: }Intersections between two regular arcs.
In this case $u$ is either a level-0 vertex of the arrangement defined by $\objectSet(\node)$
(namely when $u$ is contained in no other object $o_k\in\objectSet(\node)$), or a level-1 vertex of that
arrangement (when $u$ is contained in a single object $o_k\in\objectSet(\node)$). Note that $u$ cannot be
contained in two objects from $\objectSet(\node)$, because then $u$ would be in the interior of some
$o_k^*\in \objectSet^*(\node)$, contradicting that $u$ is a vertex of $\unionset(\node)$.
See Fig~\ref{fig:regular-irregularA}.

\textsc{Case B: }Intersections between a regular arc and an irregular arc.
Without loss of generality, assume that $u$ is the intersection of a regular arc of
$\bd o_i^*$ and an irregular arc of $\bd o_j^*$.
Note that this implies that $u$ lies in the interior of $o_j$. If there is no other object
$o_k\in\objectSet$ containing~$u$ then $u$ would be a vertex of $o_j^*$, and if there is at least one
object $o_k\in\objectSet$ containing~$u$ then $u$ would not lie on $\bd o_j^*$.
So, under the assumption that $u$ is not already a vertex of $o_j^*$, Case B does not happen.
See Fig~\ref{fig:regular-irregularB}.

\textsc{Case C:} Intersections between two irregular arcs.
In this case $u$ lies in the interior of both $o_i$ and $o_j$.
But then $u$ should also be in the interior of $o_i^*$ and $o_j^*$, so this case cannot happen.
\end{myproof}
\end{document}